# Magnetic Chains Created by Polymer-Induced Assembly of Hollow Cobalt Nanoparticles

Lin Guo*, Fang Liang, Chen Min Liu, Hui Bin Xu*, Qun Peng Zhong

School of Materials Science & Engineering, Beijing University of Aeronautics and Astronautics, Beijing, 100083, China

Xiaogang Wen, Shihe Yang*

Department of Chemistry, The Hong Kong University of Science and Technology, Clear Water Bay, Kowloon, Hong Kong, China

Wangzhi Zheng, Chinping Chen*

Department of Physics, Peking University, Beijing, 100084, China

E-mail: guolin@buaa.edu.cn

Magnetic chains of hollow cobalt nanoparticles (450-750 nm) have been synthesized by using poly(vinyl pyrrolidone) (PVP) as a template in an ethylene glycol solution. In this process, $CoCl_2 \cdot 6H_2O$ was reduced by $N_2H_4 \cdot H_2O$ in the presence of PVP. All of the Co nanoparticles are hollow with a shell of 40 nm and self-assembled into a chainlike structure that is as long as ~10 μm. At 300 K, the branched hollow Co nanoparticle chains exhibit a saturation magnetization of 37.5 emu/g, a remnant magnetization of approximately 1.55 emu/g, and a coercivity of about 66 Oe, which is more than an order of magnitude larger than that of the bulk.

In recent years, much attention has been paid to the fabrication of nano- and micro-scale hollow spheres because of their potential applications in catalysts, artificial cells, coatings, especially in delivery vehicle systems for the controlled release of drugs, cosmetics, inks, and dyes. [1–5] Above all, due to their low density, high specific surface, and large surface permeability without much sacrifice of mechanical/thermal stability, metal nanoparticles with a hollow structure exhibit a range of interesting properties superior to their solid counterparts. [6-9] For instance, the use of metals could be reduced in making conductive composites by using metallic fillers in the form of hollow nanostructures. [10] In another example, Halas et al. have demonstrated that gold nanoshells greatly expand the spectral range of surface plasmon resonance (SPR) features. [11]

Hollow metal nanostructures are often prepared by templating against existing entities, such as silica beads, [12,13] polymer beads, [14] and micelles. [15,16] Recently, transmetallation reactions between hydrophobized silver nanoparticles and hydrophobized chloroaurate and chloroplatinate ions in chloroform were reported to result in the formation of hollow gold and platinum shell nanoparticles, respectively. [17] Nano-sized Ni hollow particles were also prepared successfully via hydrothermal treatment of an alkaline solution of $Ni(DS)_2$ and $NaH_2PO_2$ at $100^0C$. [18]

Previously, we prepared a magnetic chain-like nanostructure of Ni nanoparticles using a wet chemical method in the presence of poly(vinylpyrrolidone) PVP. [19] Using a modified approach, we have successfully synthesized a similar chain-like nanostructure but consisting of Co hollow nanospheres. Here PVP is used as a dynamic template with conformational flexibility in the presence of water. To our knowledge, this is the first time a magnetic chain network of hollow metal nanoparticles is created in a simple polymer solution.

Figure 1 presents a typical X-ray diffraction (XRD) pattern of the obtained product. All the diffraction peaks can be assigned to Co hexagonal and no impurity phases such as cobalt oxide or precursor compounds have been detected, together with a hexagonal structure (JCPDS 05-0727), indicating the formation of pure cobalt with hexagonal structure. Although cobalt is usually formed in a fcc structure, [14] our reaction condition apparently favors the formation of the hexagonal phase.

Scanning electron microscopy (SEM) images of the Co products are shown in Figure 2. Clearly, the products consist of cobalt spheres ranging in size from 450 nm to 750 nm and they are in close contact with each other, forming branched necklace-like chains with a length of tens of micrometers (Figure 2a). More interesting, from the image at a higher magnification (Figure 2b), some Co spheres display broken sites and expose their hollow interiors. This is a direct evidence that the Co nanoparticles have a hollow structure with a shell thickness of ~40 nm. Although the proportion of the broken spheres appears to be only about 20% from SEM observations, the actual proportion is expected to be larger as some broken spheres may orient their holes out of sight. The partially broken hollow spheres can be more clearly seen from the high-magnification SEM image (inset of Figure 2b). The outer surfaces of these spheres are not perfectly smooth; they are somewhat coarse with a few small holes.

Transmission electron microscopy (TEM) was used to further confirm the hollow nature of the Co spheres. Some TEM images are displayed in Figure 3. From the bright field TEM image of a whole chain (Figure 3a), we see a strong contrast between the dark edges and the pale center. This confirms that all the Co spheres have a hollow with a shell thickness of 30~50 nm. The corresponding nano-beam electron diffraction (NBED) pattern (inset of Figure 3b) indicates that the hollow nanoparticle structure is polycrystalline. The interplanar spacings, obtained from the inset of Figure 3b, of 2.167, 2.023, 1.0905, 1.476, 1.248 and 1.153 nm are nicely indexed to the (100), (002), (101), (102), (110) and (103) planes of the hexagonal Co respectively, in complete agreement with the inference above from the XRD result. However, it is noticed that the NBED rings are not continuous but composed of discrete spots, a fact which suggests preferential orientations of the crystalline domains in the chains of the hollow Co nanoparticles.

More detailed information on the hollow particle chain structure was acquired using the HRTEM technique. Although the whole shell is polycrystalline, at local area, it shows preferential growth direction. Figure 4 shows the HRTEM images and corresponding FFT patterns of Co hollow particles on its outer and inner surfaces respectively. Figure 4a is from the outer surface, the interplanar spacing of 0.20 nm matches well with the distance between (002) crystal planes of hexagonal cobalt. Figure 4b is HRTEM image on the inner surface from a cracked half-sphere. Clear fringes along three symmetrical directions can be observed, their interplanar spacing of 0.22 nm can be ascribed to {100} plane of hexagonal cobalt.

Figure 5 shows the HRTEM image at the connection part of two cobalt hollow particles. The fringes with spacing of 0.22 nm (on A particle) and 0.20 nm (on B particle) can be ascribed to (100) and (002) crystal planes of cobalt, respectively.

Field-dependent magnetization was measured at 300 K by a SQUID system (Quantum Design) and the result is plotted in Figure 6. The saturation magnetization is determined as 37.5 emu/g, ~0.4 $\mu_B$ per Co atom, roughly accounting for 23% of the corresponding bulk value, ~1.7 $\mu_B$/Co. We notice that the recently reported nano-sized Ni hollow spheres also exhibit a saturation magnetization of ~24% of the bulk at 300 K. [18] Usually, for a magnetic nanoparticle, the surface spin state is amorphous-like and different from the ferromagnetic state in the core. [20] The amorphous surface magnetism would easily exhibit blocking behavior at low temperature, reflecting the surface anisotropy. At high temperature, however, the surface spins do not exhibit magnetic order due to the surface random potentials and thermal activation effects. Hence, the reduced saturation magnetization from the bulk value is expected with the geometry of a hollow nanosphere as in the case of the nano-sized Ni hollow sphere. Hysteresis loop is observed with the applied field within 1500 Oe, as shown in the inset of Figure 6. The remnant magnetization is approximately 1.55 emu/g and the remnance ratio is roughly 0.04, similar to the value determined for the Co or Fe nanowires measured with the axis perpendicular to the applied field. [21] This is reasonable since our sample was measured in a powder-collection, with the individual hollow nanoparticle chains randomly oriented. The coercivity field, $H_C$, is determined as 66 Oe, which is greatly enhanced relative to the bulk value, ~10 Oe. [22] The enhancement in the coercivity is often observed in the nano-scaled magnetic materials, attributed to the effect of surface or shape anisotropy.

We believe that the chains of the cobalt hollow spheres are formed by the interplay between complexation and reduction processes. In the beginning, $Co^{2+}$ in the solution reacts with hydrazine to form a pink complex, $[Co(N_2H_4)_3]Cl_2$, which is very stable in ambience as observed previously.[23-24] At the boiling point of the ethylene glycol, the excessive hydrazine acted as a reducing agent and converted $[Co(N_2H_4)_3]^{2+}$ to Co through homogeneous nucleation and the subsequent growth. The reactions are summarized as follows:

$Co^{2+} + 3N_2H_4 \rightarrow [Co(N_2H_4)_3]^{2+}$

$[Co(N_2H_4)_3]^{2+} + N_2H_4 \rightarrow Co\downarrow + 4NH_3\uparrow + 2N_2\uparrow + H_2\uparrow + 2H^+$

Scheme 1 illustrates a possible mechanism for the formation of the cobalt hollow nanosphere chain-like structure. It is reasonable that the linear PVP polymer chains wrap together to form random coils in the solution. [25] Such PVP pseudo-spherical coils may be tangled and joined together and they can provide adsorption sites for the precursor molecules of $[Co(N_2H_4)_3]Cl_2$. This is an important step in the synthesis of the hollow Co nanospheres. When heated to the boiling point of ethylene glycol (EG) for refluxing (~197 °C), the reduction reaction is thought to occur on the surfaces of the PVP coils in the presence of excessive hydrazine, forming a Co shell undergo the mineralization. [26] Accompanied by the formation of the Co nanoshells is the magnetic dipole-dipole attraction between them, which leads to the assembly to a chain structure. The PVP core can be removed by washing with absolute ethanol and in this way, we obtain a chain-like network of Co hollow nanospheres as observed. TEM images have demonstrated that the hollow Co nanospheres are still relatively intact even after removing the templates. Alternatively, nanoscale bubbles of gases such as water may also play a role with the assistance of PVP, because the small amount of water is found to be necessary for the formation of the hollow nanospheres. Control experiments have shown that PVP is essential to the formation of the cobalt hollow nanosphere chain-like structure. In the absence of PVP, only solid Co microspheres were obtained without the chain-like structure.

In closing, we would like to emphasize that this is the first time a hollow Co nanoparticle chain-like structure is created by a simple polymer solution method. This method has the advantages of simplicity, high yield, and mild reaction conditions. Therefore, it represents an attractive path to large-scale assembly of metallic hollow nanospheres. We have revealed a significantly enhanced (x7) magnetic coercivity at 300 K for the Co hollow nanospheres from the bulk value. The excellent stability, the hollow nanostructure, and the hierarchical self-assembled architecture afford a model system for fundamental investigations and promising applications in various fields of nanotechnology. We envisage that this simple method is general and can be applied to the synthesis of hollow nanosphere chain-like structures of many other metals.

Experimental

All chemicals (Acros) used in this experiment were analytical grade and used without further purification. The growth of the pure cobalt hollow particles chain structure was carried out in a solution phase system. First, 0.3032 g $CoCl_2 \cdot 6H_2O$ and 0.4260 g PVP (MW 58,000) were dissolved in 30 ml ethylene glycol (EG) by intensive stirring for 2 hours and a homogenous transparent mauve solution was obtained. 1.5 ml hydrazine monohydrate (50% Vol. A.R.) was added dropwise to the well-stirred mixture at room temperature by simultaneous vigorous agitation. The solution turned turbid and pink. The stirring process lasted for at lease 1 hour after finishing the dropping in order for the reaction to go completely. The mixture was subsequently heated to the boiling point of ethylene glycol (EG) for refluxing (~197 $^oC$). Numerous bubbles were formed in the flask with niffy smell. After refluxing for 4 hours, the color of the solution turns from pink to dark and the dark precipitate could be achieved. Centrifugation was used to separate the precipitate, which was rinsed with absolute ethanol for 6 times. Subsequently, the volatile solvent was evaporated in vacuum at 80 $^oC$ and finally a loose dark powder was obtained. These as-prepared products were used for characterization.

X-ray powder diffraction (XRD, Rigaku, Dmax2200, Cu-kα) was used for the structural determination. Further microstructural analyses were performed using a scanning electron microscopy (SEM) (JSM-5800 with accelerating voltage of 15 kV) and analytical transmission electron microscopy (TEM) (JEOL 2100F). TEM samples were prepared by dispersing the powder products in alcohol by ultrasonic treatment, dropping the suspension onto a holey carbon film supported on a copper grid, and drying it in air. Magnetic properties of the sample were measured using a Quantum Design SQUID magnetometer.

Acknowledgements

This project was financially supported by National Natural Science Foundation of China and Program for New Century Excellent Talents in University (NCET) as well as by Engineering Research Institute, Peking University (ERIPKU). SY acknowledges the support from the Research Grants Council of Hong Kong.

**Figures.**

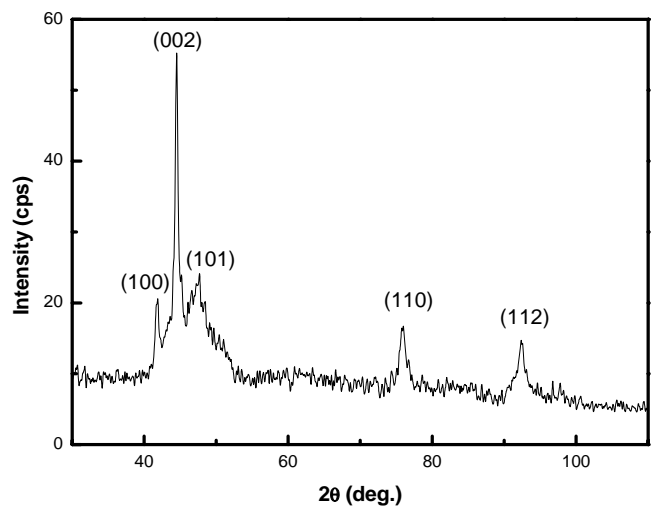

Figure 1 XRD pattern of the as-synthesized cobalt nanoparticle chain network. Note that all of the peaks can be indexed to hexagonal cobalt.

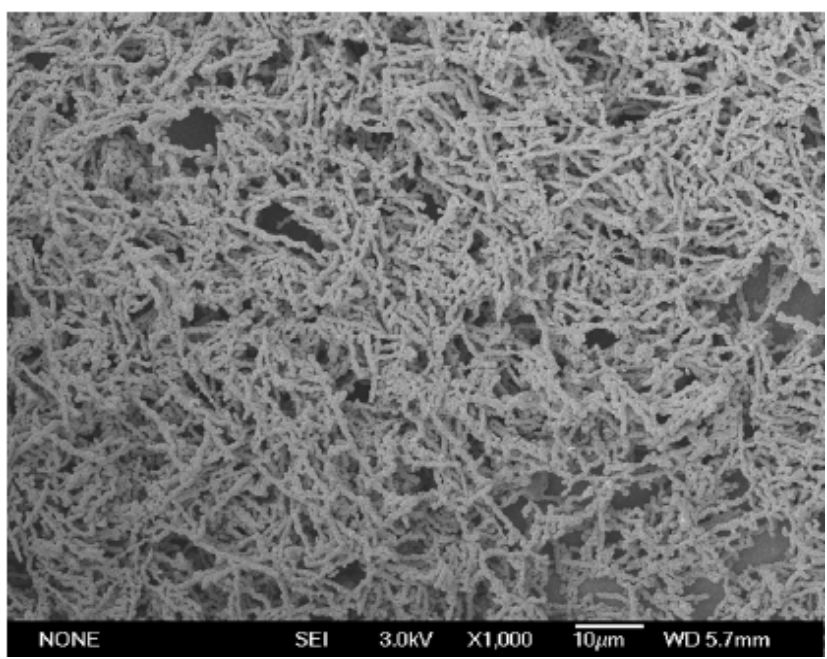

Figure 2a SEM image of the chainlike structure of cobalt hollow spheres at a low magnification.

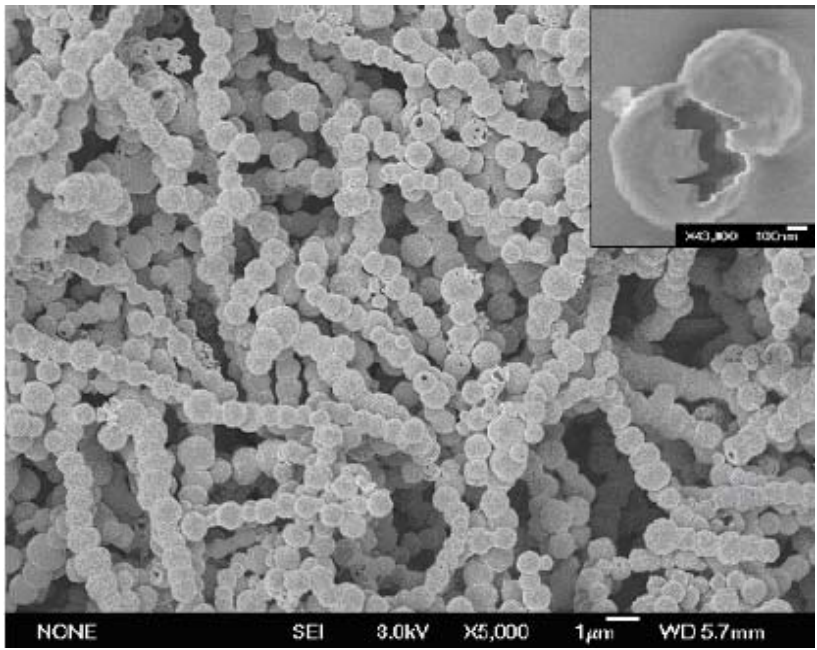

Figure 2b SEM image at a higher magnification (inset: SEM image of two broken hollow sphere).

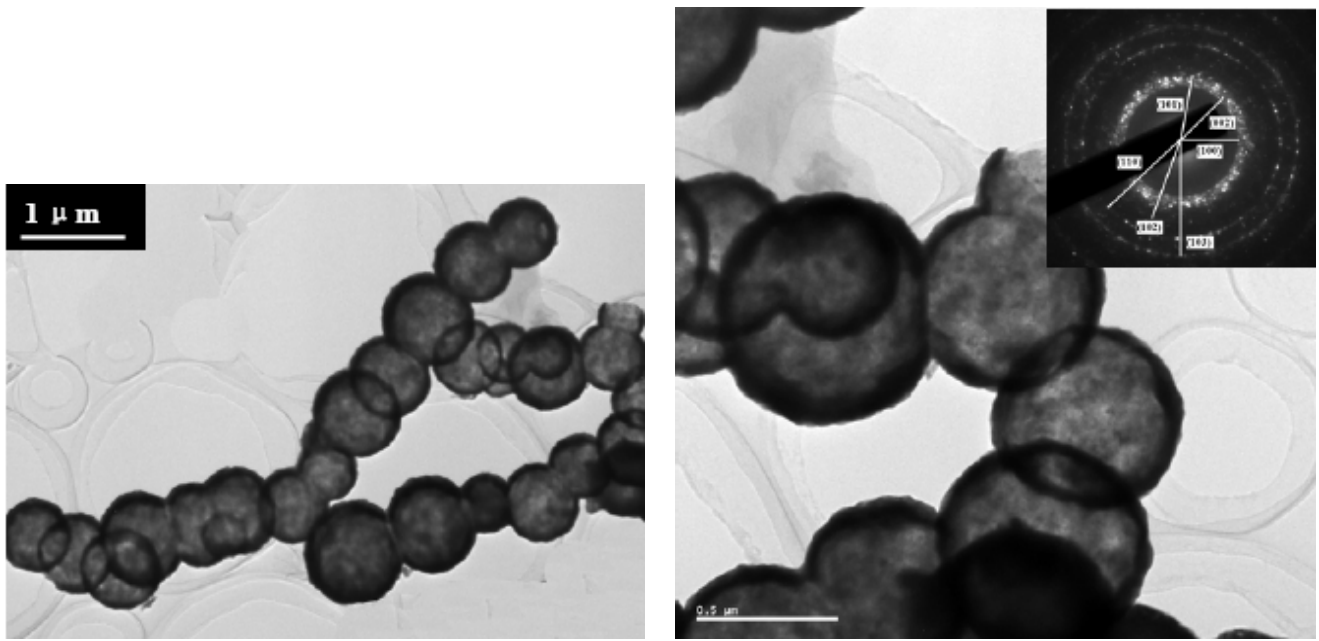

Figure 3a   TEM image of a whole chainlike structure

Figure 3b Bright field TEM image of the chainlike hollow structure at a higher magnification. The insert shows the corresponding nano-beam electron diffraction (NBED) pattern.

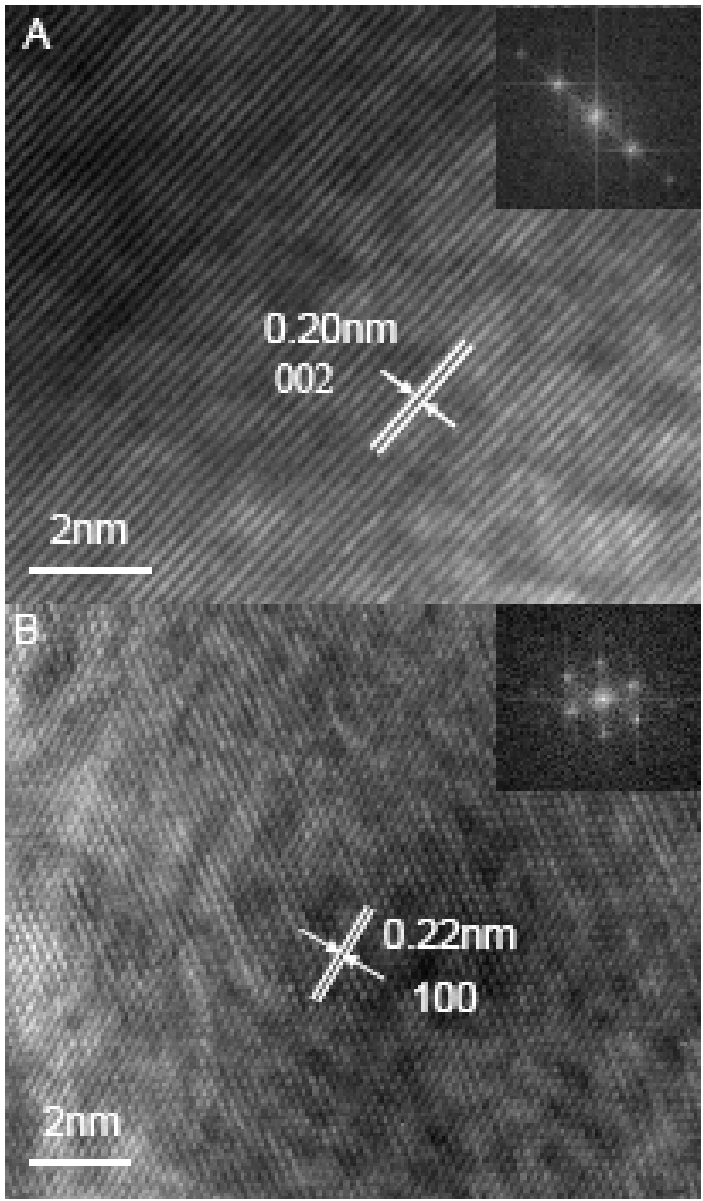

Figure 4  Lattice resolved HRTEM image taken at the edge between the hollow and the shell of a hollow sphere.

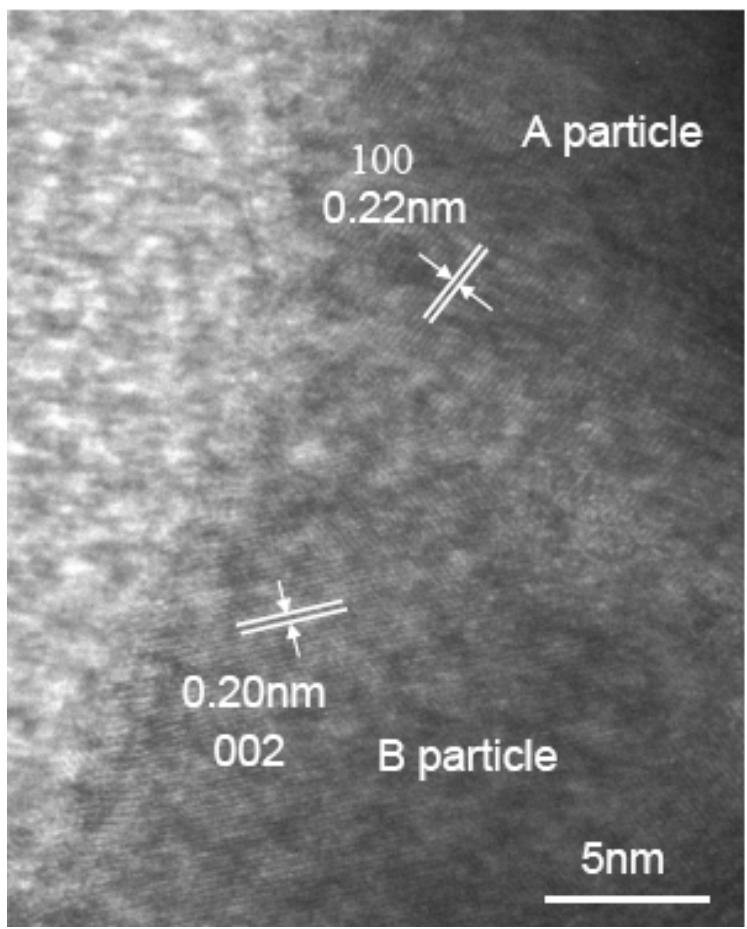

Figure 5 HRTEM image at the connection part of two cobalt hollow particles.

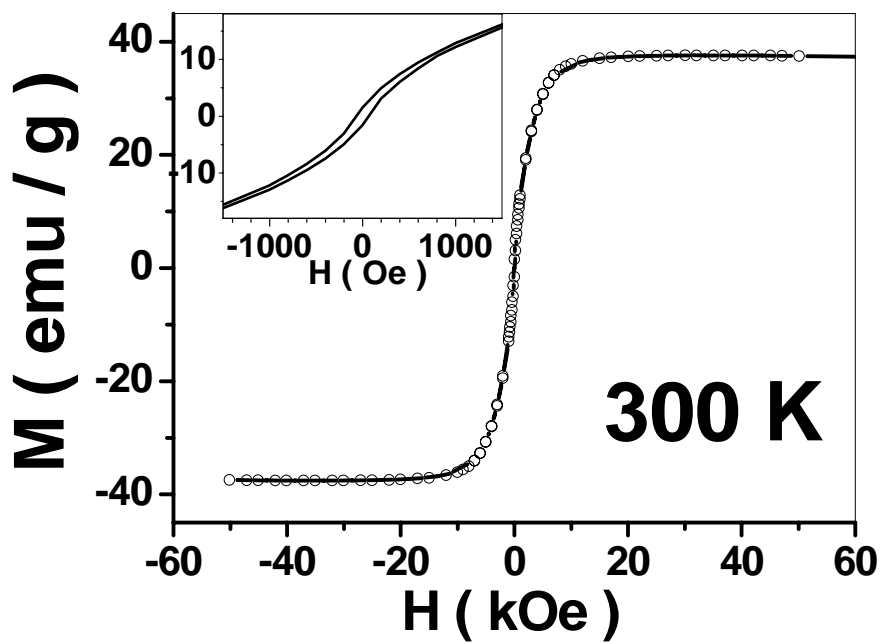

Figure 6  Magnetization versus field at 300 K. The saturation magnetization is 37.5 emu/g (~ 0.4 $\mu_B$ per Co atom). The inset shows the hysteresis loop at low field, below 1500 Oe. The coercivity is determined as 66 Oe, with a remnant of about 1.55 emu/g.

**Scheme 1.  Illustration of the formation of the cobalt hollow sphere chainlike structure.**

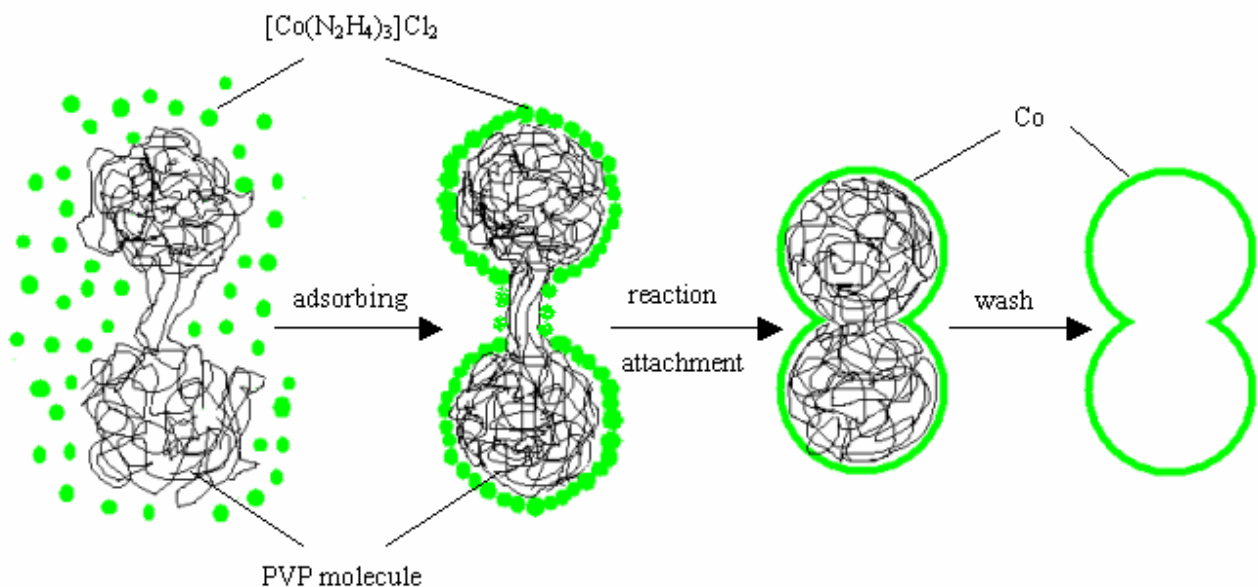